\documentclass[10pt,letterpaper]{article}
\usepackage{opex3}
\usepackage{subfigure}
\usepackage{amsmath}
\usepackage{graphicx}
\usepackage{epstopdf}
\usepackage{cite}
\newcommand{\Der}{\mathrm{d}}

\newcommand{\Eul}{\mathrm{e}}

\begin{document}

%%%%%%%%%%%%%%%%%% title page information %%%%%%%%%%%%%%%%%%
\title{Incoherent on-off keying with classical and~non-classical light}

\author{Marcin Jarzyna, Piotr Kuszaj, and Konrad Banaszek$^*$}

\address{Faculty of Physics, University of Warsaw, Pasteura 5, 02-093 Warsaw, Poland}

\email{$^{*}$Konrad.Banaszek@fuw.edu.pl} %% email address is required



\begin{abstract}
We analyze the performance of on-off keying (OOK) and its restricted version pulse position modulation (PPM) over a lossy narrowband optical channel under the constraint of a low average photon number, when direct detection is used at the output. An analytical approximation for the maximum PPM transmission rate is derived, quantifying the effects of photon statistics on the communication efficiency in terms of the $g^{(2)}$ second-order intensity correlation function of the light source. Enhancement attainable through the use of sub-Poissonian light is discussed.
\end{abstract}

 % REPLACE WITH CORRECT OCIS CODES FOR YOUR ARTICLE

%%%%%%%%%%%%%%%%%%%%%%% References %%%%%%%%%%%%%%%%%%%%%%%%%

%%%\section{Introduction}

Optical transmission has become the backbone of modern high-throughput information infrastructures. It is also studied in the context of more specialized applications, such as deep-space communication. Transfer rates attainable over optical channels are fundamentally limited by the available signal power, spectral range, and noise effects in the medium transmitting signals, which altogether determine the ultimate channel capacity implied by the laws of quantum physics \cite{Caves,Shapiro,GiovGuhaPRL2004,GiovGarcXXX13}.

One of the basic models of optical communication is a lossy narrowband channel, where a single bosonic mode transmitted in each use undergoes linear amplitude damping. A natural physical constraint in this scenario is a bound on the average power. Among common modulation schemes, on-off keying (OOK) and its restricted version pulse position modulation (PPM), which use just two states: an empty vacuum bin and a non-zero light pulse \cite{GuhaHabiJMO2011,WaseSasaJOCN2011,AntoMecoJLT2014}, are known to approach the capacity of the narrowband channel at very low average energy levels even if incoherent, direct photodetection is used \cite{KochWang2014}. In this paper, we present a systematic study of the effects of the photon statistics on the transmission rate in the incoherent scenario with direct detection. Using the quantum theory of photodetection \cite{KellKleiPRA1964}, we identify the well-known $g^{(2)}$ normalized intensity correlation function at zero delay \cite{GlauPR63} as the relevant characteristics that explicitly enters the expression for Shannon mutual information. The analysis is extended into the non-classical region of sub-Poissonian photon statistics, quantifying how much the transmission rate can be enhanced by the use of non-classical states of light.

%%%\section{Transmission rate}

\begin{figure}
\centering
\includegraphics[angle=0,width=0.92\textwidth]{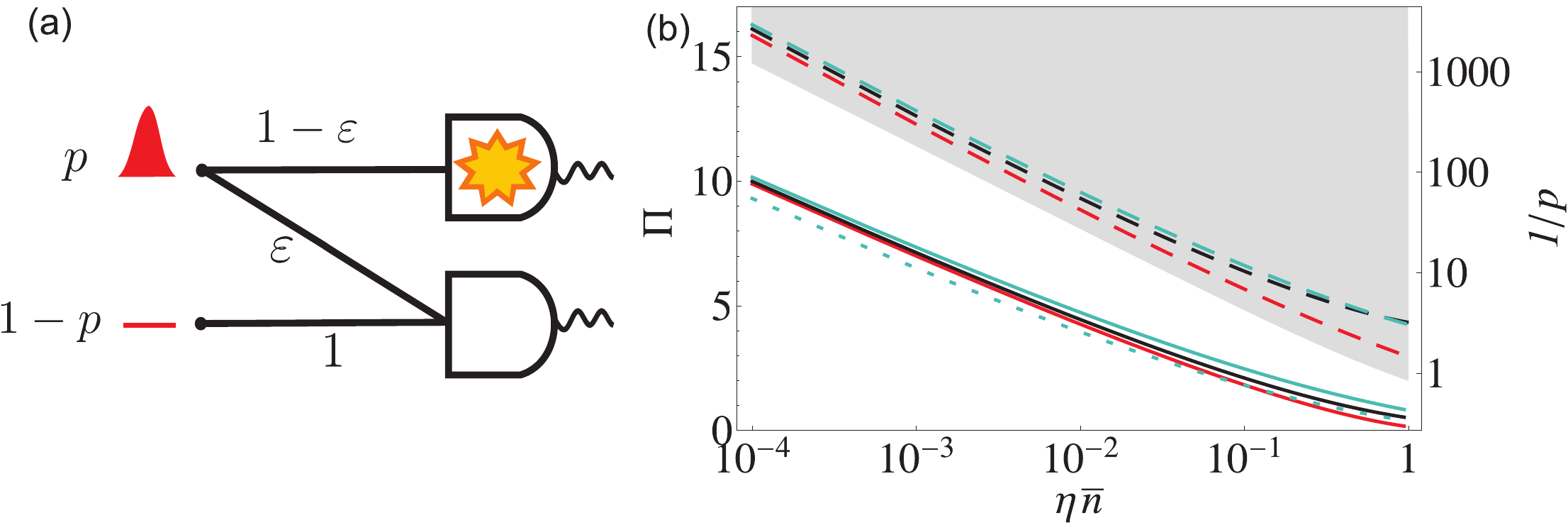}
\caption{(a) OOK with direct detection represented as a binary asymmetric channel. For the description of symbols, see the main text. (b) Left scale: the approximate analytical expression for photon information efficiency $\Pi$ derived in Eq.~(\ref{Eq:PIEApprox}) as a function of the average output photon number $\eta\bar{n}$ (red solid line) compared with the result of numerical optimization assuming Poissonian photon statistics for PPM (black solid line) and OOK (blue solid line) schemes. The blue dotted line depicts optimized PIE for OOK with Poisonian statistics and dark count probability $0.25\eta\bar{n}$ per time bin. The grey region represents values of PIE beyond the capacity limit of a single-mode bosonic channel, equal to $\log_2(1/\eta\bar{n}) + (1+1/\eta\bar{n})\log_2(1+\eta\bar{n})$. Right scale: the optimal value of $1/p$ maximizing mutual information given by the approximate expression in Eq.~(\ref{Eq:IPPMapprox}) (red dashed line) compared to numerical results for PPM (black dashed line) and OOK (red dashed line) with Poissonian photon statistics.}
\label{Fig:DiagramPIEs}
\end{figure}

From the information-theoretic viewpoint, OOK with direct detection is described by a binary asymmetric channel depicted in Fig.~\ref{Fig:DiagramPIEs}(a). A non-zero pulse and an empty bin are prepared with respective probabilities $p$ and $1-p$. Neglecting detector dark counts, empty bins always generate no-click events. Pulses are detected with a probability $1-\epsilon$. The crossover probability $\epsilon$ that a pulse will not generate a detector click
%, providing the outcome identical with that for an empty bin,
depends on the pulse energy, channel transmission, and detection efficiency, but more specifically also on the actual photon statistics of the input pulse, which will be the focus of this analysis. Our objective will be to optimize the attainable transmission rate under the constraint of a fixed average energy per channel use.

In order to gain an intuitive insight into the performance of OOK, we will first consider its restricted version known as PPM, where exactly one non-zero pulse is prepared in a sequence of otherwise empty bins of length $1/p$. Although in this scenario $p$ is implicitly assumed to be the inverse of an integer, it can be approximately treated as a continuous real parameter for long bin sequences. If direct detection is used at the output, PPM is described by a $1/p$-ary erasure channel \cite{CoverThomas}, where $1/p$ words are defined by the position of the non-zero pulse within a sequence, with the erasure probability equal to $\epsilon$. Consequently, the Shannon mutual information, which characterizes how much knowledge about inputs can be recovered from detection events at the output, renormalized to a single time bin reads $I_{\text{PPM}} = p (1-\epsilon) \log_2 (1/p)$, where the logarithm is taken to base 2 for information expressed in bits. This quantity defines the maximum attainable transmission rate in the communication scheme under consideration \cite{CoverThomas}.

Under the constraint of low average energy per channel use we expect that the detection probability $1-\epsilon$ of the non-zero pulse will be small.
According to quantum theory of photodetection \cite{KellKleiPRA1964}, which covers also the use of non-classical states of light, the probability of a no-count event is expressed by the detector efficiency $\eta$ and the photon number operator $\hat{n}$  for the incident pulse as $\epsilon = \langle : e^{-\eta \hat{n}} : \rangle$, where $:\ldots:$ denotes normal ordering of field creation and annihilation operators and the angular brackets stand for the quantum mechanical expectation value over the light pulse state. In the communication scenario discussed here, the parameter $\eta$ can be interpreted as the non-unit transmission of the optical channel. Our analysis will rely on the expansion of the crossover probability up to the second order in the exponent $\eta\hat{n}$:
\begin{equation}
\label{Eq:ClickProb}
 \epsilon  \approx 1 - \eta \langle \hat{n} \rangle + \frac{1}{2} \eta^2  \langle \, : \! \hat{n}^2 \! :\, \rangle
 = 1 - \eta \mu + \frac{1}{2} g^{(2)} \eta^2 \mu^2.
\end{equation}
In the latter expression we denoted as $\mu = \langle \hat{n} \rangle$ the mean photon number in the non-zero pulse and recalled the standard definition of the normalized second-order intensity correlation function at zero delay $g^{(2)} = \langle \, : \! \hat{n}^2 \! :\, \rangle / \langle \hat{n} \rangle^2$ \cite{GlauPR63,Schmunk2012}.
The approximation introduced in Eq.~(\ref{Eq:ClickProb}) is valid when the non-zero pulse at the channel output has sufficiently low mean photon number $\eta\mu \ll 1$ and intensity correlation functions of the order higher than two are not extravagantly large. Furthermore, the expansion applied in Eq.~(\ref{Eq:ClickProb}) is exact when the input pulse is prepared as a non-classical mixture of Fock states containing up to two photons.

%%%%\section{Optimization}

We will now optimize mutual information under the constraint of given average power, which for a narrowband channel can be expressed in terms of the average photon number per bin, denoted as $\bar{n} = p\mu$. This
%links the mean photon number for the non-zero pulse $\mu$ with the probability $p$ of its preparation:
%\begin{equation}
%\label{Eq:Constraint}
%\bar{n} = p\mu.
%\end{equation}
%The above
relation allows us to express $p$ in terms of $\mu$ and select the latter as the free parameter when optimizing mutual information, which after applying expansion (\ref{Eq:ClickProb}) takes the form
\begin{equation}
\label{Eq:IPPMapprox}
I_{\text{PPM}} = \eta\bar{n} \left( 1- \frac{1}{2} g^{(2)} \eta\mu \right) \log_2 \frac{\mu}{\bar{n}}.
\end{equation}

We will assume for now that the intensity correlation function $g^{(2)}$ is a fixed property of the light source. Let us note that this characteristics remains constant when amplitude modulation is used to adjust the pulse energy. For classical light, when $g^{(2)} \ge 1$, the mean photon number in the non-zero pulse can take any non-negative value. Its optimal value in the case of PPM encoding can be found explicitly for the expression given in Eq.~(\ref{Eq:IPPMapprox}) %%%and the constraint (\ref{Eq:Constraint})
from the derivative  $\Der I_{\text{PPM}}/\Der\mu =0$. The solution is given in terms of the Lambert $W$ function \cite{LambertW} as
\begin{equation}
\label{Eq:muopt}
\mu^{\text{clas}} = \frac{2}{\eta g^{(2)}}\left[  W\left( \frac{2e}{g^{(2)} \eta\bar{n}} \right) \right]^{-1}.
\end{equation}
For large arguments of the Lambert function we can use its asymptotic form $W(x) \approx \ln x $. This implies that in the regime of a low average photon number at the output, when $\eta\bar{n} \ll 1$, we have a hierarchy $ \eta\bar{n} \ll \eta\mu^{\text{clas}} \ll 1$, which justifies the application of Eq.~(\ref{Eq:ClickProb}) and the treatment of the PPM sequence length as a continuous parameter. The optimal value of mutual information $I_{\text{PPM}}^{\text{clas}}$ can be written as a product of $\eta\bar{n}$ and the quantity known as photon information efficiency (PIE), i.e.\ the average amount of information transmitted per output photon:
\begin{equation}
\label{Eq:Ippmopt}
I_{\text{PPM}}^{\text{clas}} = \eta\bar{n} \Pi \bigl(g^{(2)}\eta\bar{n} \bigr)
\end{equation}
where the function $\Pi$ is given explicitly by
\begin{equation}
\label{Eq:PIEApprox}
\Pi(\eta\bar{n}) = \left\{ \left[ W\left( \frac{2e}{\eta\bar{n}} \right) \right]^{-1} -1 \right\}
\log_2 \left[\frac{\eta\bar{n}}{2} W \left( \frac{2e}{\eta\bar{n}} \right) \right].
\end{equation}
This formula provides an analytical expression for PIE in the case of PPM assuming Poissonian pulse statistics with $g^{(2)}=1$, derived under the approximation $(\ref{Eq:ClickProb})$ when the average output photon number per bin is fixed at $\eta\bar{n}$. Approximating in Eq.~(\ref{Eq:PIEApprox}) the Lambert $W$ function by the natural logarithm provides the asymptotic form of Poissonian PIE for $\eta\bar{n} \ll 1$ discussed in~\cite{KochWang2014}.

Figure~\ref{Fig:DiagramPIEs}(b) compares the analytical expression for $\Pi(\eta\bar{n})$ derived in Eq.~(\ref{Eq:PIEApprox}) with direct numerical optimization of PIE for PPM and OOK schemes when the crossover probability is given by the Poissonian statistics without any approximation, $\epsilon = e^{-\eta\mu}$. Because in the case of Poisson distribution the only effect of non-unit channel transmission is lower mean value of the photon statistics at the output, the results depend only on the product $\eta\bar{n}$. The optimal values of $1/p=\mu/\bar{n}$ depicted in Fig.~\ref{Fig:DiagramPIEs}(b) confirm that in the regime $\eta\bar{n} \ll 1$ the number of bins in the PPM encoding can be treated as a continuous parameter. For reference, we also display the ultimate quantum limit on PIE imposed by the capacity of a single-mode bosonic channel \cite{GiovGuhaPRL2004,GiovGarcXXX13}. Figure~\ref{Fig:DiagramPIEs}(b) shows that PIE is a monotonically decreasing function of its argument. This fact allows us to discuss the  case of general photon statistics covered by the formula derived in Eq.~(\ref{Eq:Ippmopt}). According to this expression, the intensity correlation function $g^{(2)}$ simply rescales the average output photon number $\eta\bar{n}$ as the argument of $\Pi$. Therefore excess classical noise in photon statistics, when $g^{(2)} > 1$, decreases mutual information. The effective PIE is equal to the value obtained for the average photon number $g^{(2)}\eta\bar{n}$ in the Poissonian scenario.

%%%%\section{Non-classical light}

%%%Extending the argument presented in the preceding section
The simple scaling law of PIE with the second-order intensity correlation function $g^{(2)}$
suggests that using sub-Poissonian light characterized by $g^{(2)}<1$ can increase Shannon information beyond that attainable with classical light for fixed $\eta$ and $\bar{n}$. In the lossless case this effect can be related to a simple observation that the use of Fock states allows one in principle to suppress the crossover probability $\epsilon$. However, with increasing channel losses Fock states produce output photon statistics with a large vacuum contribution, which should limit the attainable enhancement despite non-classical $g^{(2)}$. Mathematically, this limitation stems from the fact that the optimal mean photon number found in Eq.~(\ref{Eq:muopt}) may be unphysical for $g^{(2)}<1$. This can be seen by considering the non-negativity of the photon number variance for the non-zero pulse, $\langle \hat{n}^2 \rangle - \langle \hat{n} \rangle^2 \ge 0$. Expressing the expectation value of the squared photon number operator as $\langle \hat{n}^2 \rangle = \langle \, : \! \hat{n}^2 \! :\, \rangle + \langle \hat{n} \rangle = g^{(2)} \mu^2 + \mu$ after simple algebra yields a bound
\begin{equation}
\label{Eq:mug2constraint}
\mu \le \frac{1}{1-g^{(2)}}.
\end{equation}
One should note that the above inequality can be saturated only when $g^{(2)} = 1-1/m$ with an integer $m$ corresponding to a pure $m$-photon Fock state which guarantees vanishing photon number variance. For other values of $g^{(2)}$ the minimum photon number variance is reached for a mixture of two consecutive Fock states. Therefore a tight bound on $\mu$ is obtained from the inequality $\langle \hat{n}^2 \rangle - \langle \hat{n} \rangle^2 \ge (\langle \hat{n} \rangle - \lfloor \langle \hat{n} \rangle \rfloor)(1 - \langle \hat{n} \rangle + \lfloor \langle \hat{n} \rangle \rfloor)$, where $\lfloor \ldots \rfloor$ denotes the integer part. However, the more relaxed bound (\ref{Eq:mug2constraint}) will lead us to a simple analytical formula approximating mutual information.

\begin{figure}
\centering
\includegraphics[angle=0,width=0.96\textwidth]{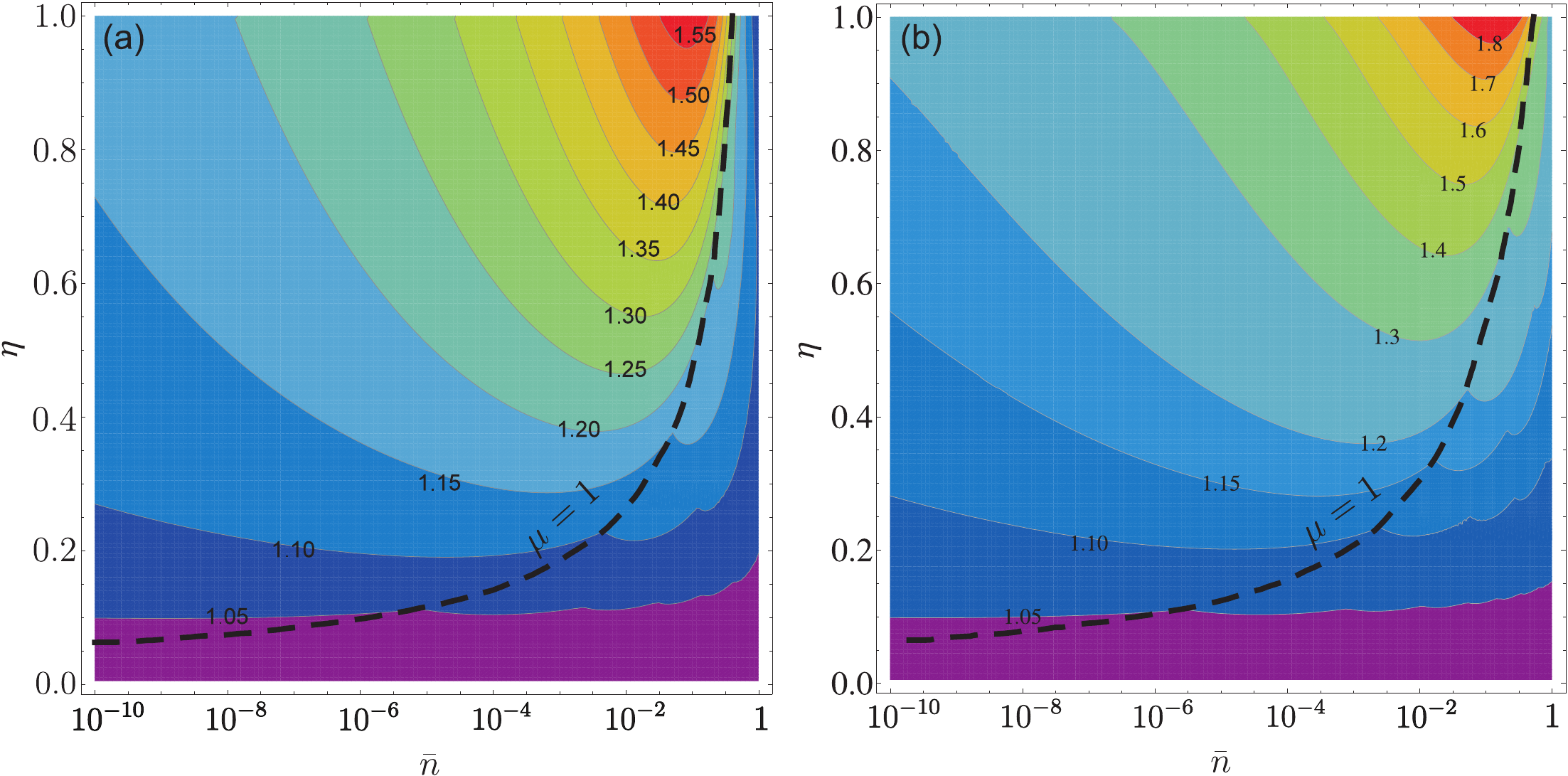}
\caption{Ratio of mutual information optimized over non-classical two-component mixtures of adjacent Fock states to that attainable with Poissonian statistics of the non-zero pulse for (a) PPM and (b) OOK schemes. The dashed lines indicate regions where single-photon states with $\mu=1$ are optimal in the non-classical case.}
\label{Fig:Ratio}
\end{figure}

Let us now discuss the problem of optimizing mutual information $I_{\text{PPM}}$ given in Eq.~(\ref{Eq:IPPMapprox}) with respect to both $\mu$ and $g^{(2)}$ for non-classical light, taking into account the physical constraint (\ref{Eq:mug2constraint}).
Using the fact that for a fixed $g^{(2)}$ the expression on the right hand side of Eq.~(\ref{Eq:IPPMapprox}) has a single maximum in $\mu$, whose value grows with decreasing $g^{(2)}$, it is easy to show that %%% in the non-classical region $0 \le g^{(2)} < 1$
optimal information is attained on the boundary of the region defined by the physical constraint, where the inequality (\ref{Eq:mug2constraint}) is saturated. Inverting this relation yields $g^{(2)} = 1-1/\mu$, which inserted into Eq.~(\ref{Eq:IPPMapprox}) reduces the problem to single-parameter optimization of $I_{\text{PPM}}$ over $\mu \ge 1$. The result can be written in an analytical form as one of two cases:
\begin{equation}
\label{Eq:IPPPopt}
I_{\text{PPM}}^{\text{opt}} =
\begin{cases}
\eta \bar{n} \log_2 \frac{1}{\bar{n}} & \text{if $\eta \ge 2/\ln \frac{1}{\bar{n}}$,} \\
\eta \bar{n} \left( 1+ \frac{\eta}{2} \right) \Pi \left( \frac{\eta\bar{n}}{1+\eta/2} \right) & \text{if $\eta < 2/\ln \frac{1}{\bar{n}}$.}
\end{cases}
\end{equation}
In the first case, mutual information is maximized by the use of one-photon Fock states, $\mu=1$. This holds for the channel transmission $\eta$ above the threshold value $2/\ln ({1}/{\bar{n}})$ which tends to zero with $\bar{n} \rightarrow 0$, but with a very slow scaling in $\bar{n}$. In this regime the expansion applied in Eq.~(\ref{Eq:ClickProb}) is obviously exact, with $g^{(2)} = 0$. It can be easily verified that for the second case in Eq.~(\ref{Eq:IPPPopt}), in the regime $\eta\bar{n} \ll 1$ the optimal $\mu$ satisfies $\eta\mu \ll 1$, confirming the consistency of using the approximate expression for the crossover probability $\epsilon$. The corresponding expression for maximum mutual information is enhanced by a multiplicative factor $1+ \eta/2$, appearing also in the argument of the function $\Pi$. This factor approaches one with increasing channel losses, which limits the non-classical enhancement.

The approximate analytical results presented above provide guidance in numerical calculations based on complete expressions for mutual information and the crossover probability. As the indication of the transmission rate attainable with classical light we use mutual information optimized over the mean photon number $\mu$ in the non-zero pulse assuming the exact Poissonian expression for the no-count event probability $\epsilon = \Eul^{-\eta\mu}$. %%%The maximum information depends only on the average photon number at the channel output given by the product $\eta\bar{n}$.
The results of optimization have been presented in Fig.~\ref{Fig:DiagramPIEs}(b) in the form of respective PIEs for OOK and PPM schemes, which need to be multiplied by $\eta\bar{n}$ to obtain mutual information.

In the non-classical regime we performed optimization over a one-parameter family of states characterized by the mean photon number $\mu$. Its actual form is motivated by the physical constraint derived in Eq.~(\ref{Eq:mug2constraint}). For an integer $\mu$ we assume a $\mu$-photon Fock state, whereas for a non-integer $\mu$ we take a mixture of adjacent Fock states containing $\lfloor \mu \rfloor$ and $\lfloor \mu \rfloor+1$ photons with respective weights $1-\mu+\lfloor \mu \rfloor$ and $\mu-\lfloor \mu \rfloor$. The no-count probability is given in this case by $\epsilon = [1-\eta(\mu - \lfloor \mu \rfloor)] (1-\eta)^{\lfloor \mu \rfloor}$. We found that in the case of PPM the relative difference between  numerical results and the approximate formula derived in Eq.~(\ref{Eq:IPPPopt}) is less than $2.5\%$ for $\bar{n} \le 0.1$ over the full range of channel transmission $0 \le \eta \le 1$.

Figure~\ref{Fig:Ratio} depicts the enhancement possible by the application of non-classical states of light, quantified as the ratio of mutual information optimized over the family of two-component Fock state mixtures to that attainable using Poissonian statistics. A noticeable improvement can be achieved only for low channel losses and relatively high average input photon number $\bar{n}$. Further, dashed lines in Fig.~\ref{Fig:Ratio} bound the regions where Shannon information is maximized by the use of single photon Fock states. The  results for PPM and OOK schemes are qualitatively similar, although the latter gives a slightly higher enhancement. In either case, improvement of the transmission rate by at least 10\% is possible for channel transmission $\eta$ above $20\%$.

%%%%\section{Conclusions}

In conclusion, we have derived an approximate analytical form of optimized Shannon information for PPM encoding and direct detection under the constraint of a low average energy per bin. The obtained expression quantifies the role of excess noise in photon statistics in terms of the normalized second-order intensity correlation function. Using single photons can enhance the attainable transmission rate through the reduction of the cross-over probability, but this effect becomes substantial only in the high-transmission regime.
In addition to channel losses, another common imperfection in optical communication is related to detector dark counts. A preliminary analysis suggests that this does not compromise the operation of the schemes considered here, as long as the dark count rate remains at a sufficiently low level compared to the average output signal power. As an illustration, in Fig.~\ref{Fig:DiagramPIEs}(b) we depict PIE for OOK with Poissonian statistics and dark count probability $0.25\eta\bar{n}$ per time bin.

From a more general perspective, results presented in this paper highlight the role of optical coherence in achieving the capacity of a single-mode bosonic channel. In the lossless case, channel capacity given by the Holevo quantity can be saturated either with a Gaussian ensemble of coherent states, or by using a set of phase-invariant Fock states taken with Bose-Einstein distribution \cite{Caves}. For a lossy channel, coherent states remain pure at the channel output yielding maximum Holevo quantity \cite{GiovGuhaPRL2004,GiovGarcXXX13}, whereas Fock states are transformed into statistical mixtures, which lowers attainable information. First coherent receivers offering transmission rates beyond direct detection have been demonstrated in proof-of-principle settings only recently \cite{TsujFukuOE2010,ChenHabiNPH2012}. This however leads to questions about effects of phase fluctuations and local oscillator noise to which coherent detection schemes may be overly sensitive \cite{OlivaCialPRA2013,JarzBanaJPA2004}. It is therefore important to compare the performance of practical communication schemes under different constraints on available resources \cite{TakeGuhaPRA2014}.

%%It should be therefore useful to consider as a benchmark completely dephased communication schemes with general quantum states of light, such as scenarios discussed here.

\section*{Acknowledgments}
We thank R. Demkowicz-Dobrza\'{n}ski, F. Grosshans, and M. G. Raymer for insightful discussions. This research was partly supported by the EU 7th Framework Programme projects SIQS (Grant Agreement No. 600645) and PhoQuS@UW (Grant Agreement No. 316244).

\end{document}